\documentclass[amssymb,floatfix,graphicx,reprint,authoryear,12pt]{elsarticle}

\usepackage{amsmath}
\usepackage{dcolumn}% Align table columns on decimal point
\usepackage{bm}% bold math
\usepackage[mathlines]{lineno}% Enable numbering of text and display math
\usepackage[utf8]{inputenc}
\usepackage[english]{babel}
\usepackage{mathptmx}
\usepackage{caption}
\usepackage{subcaption}
\usepackage{setspace}
\usepackage{color}
\usepackage{float}

\journal{International Journal of Multiphase Flow}

\begin{document}

\begin{frontmatter}

\title{Dynamics of Evaporating Respiratory Droplets in the Vicinity of Vortex Dipoles}

\author[add1]{Orr Avni}

\author[add1]{Yuval Dagan}
\ead{yuvalda@technion.ac.il}

\address[add1]{Faculty of Aerospace Engineering, Technion - Israel Institute of Technology, Haifa, 320003, Israel.}

\begin{abstract}
A new mathematical analysis of exhaled respiratory droplet dynamics and settling distances in the vicinity of vortical environments is presented.
Recent experimental and theoretical studies suggest that vortical flow structures may enhance the settling distances of exhaled respiratory droplets beyond the two-meter distancing rule recommended by health authorities lately.
We propose a mathematical framework to study the underlying physical mechanism responsible for the entrapment and subsequently delayed settling times of evaporating droplets and solid particles.
A dipolar vortex is considered self-propelling through a cloud of micron-sized evaporating droplets.
This configuration might be utilized to approximate an indoor environment in which similar unsteady vortical flow structures interact with exhaled respiratory droplets. 
We demonstrate the vortex dipole effect on droplet and solid particles settling distances, depending on the evaporation rate, the vorticity of the dipole, and the droplet's initial diameter and location relative to the vortex core. 
Our theoretical analysis reveals non-intuitive interactions between the vortex dipole, droplet relaxation time, gravity, and mass transfer.
The existence of optimal conditions for maximum displacement is suggested, where the droplet entrainment reaches up to an order of magnitude larger than the vortex core length scale.
We present a basic model that may be applied for evaluating the spread of exhaled respiratory droplets in vortical environments. Our theoretical study suggests that exhaled respiratory droplets initially at rest can translate to significant distances, hence implying that vortical flow might enhance the transmission of airborne pathogens.
\end{abstract}


\begin{keyword}
drops and bubbles \sep vortex flows \sep multiphase and particle-laden flows
\end{keyword}

\end{frontmatter}

Respiratory activities of an infected individual, such as talking, coughing, or sneezing, lead to the release of variable amounts of virus-laden droplets. Exposure to contaminated droplets might infect nearby individuals, transmitting the disease and increasing the risk of it spreading rapidly and uncontrollably~\citep{world2020transmission}. 
Possible routes of transmission that have been widely recognized are the "airborne droplet" and "airborne aerosol" routes~\citep{Tellier2019,Li2021basic,chagla2020re}. The first involves the deposition of larger droplets on the mucosal surfaces (e.g., eyes, nose, mouth), while the second refers to the inhalation of finer droplets by the susceptible person.
Thus, the persistence and dispersion of sub-micron-sized aerosols are of great concern, especially due to implications regarding public gathering in closed environments, although the extent to which airborne mechanisms are responsible for the vast COVID-19 outbreak remains elusive~\citep{asadi2020coronavirus, mittal2020mathematical,smith2020aerosol,bourouiba2020turbulent,Chen2020}.

Ventilation plays a key role in preventing aerosol transmission in close environments, as it dictates the rate at which the virus-laden air is replaced with fresh air, reducing the concentration of contaminated droplets in the room and decreasing the infection probability~\citep{Morawska2020,Li2007,Sundell2021}.
Thus, better ventilation of poorly ventilated public environments that have been suspected as potential sites of transmission, such as hospital elevators and public transport, was suggested as a preventive measure, assisting in reducing the spread of SARS-CoV-2~\citep{VanRijn2020,Qian2021,Somsen2020}.
However, the turbulent air created by various natural and fan-driven ventilation is also known to be effective in the transport of airborne pathogens and might propel the virus-laden droplet to considerable distances~\citep{Li2005,Li2007}.

Past and recent studies on the hydrodynamics of exhaled respiratory droplets show that for most scenarios in which human coughing and sneezing are involved, the settling distances of largest-scale droplets do not exceed $1.5-2m$ ~\citep{Xie2007,Wang2020}, and hence the 2-meter rule for social distancing recommended by most countries during the outbreak of COVID-19.
However, most research efforts in hydrodynamic dispersion of exhaled respiratory droplets focus on either speaking, coughing, or sneezing into a quiescent environment~\citep{bourouiba2014violent,Balachandar2020,Aiyer2020,Fabregat2021}, or assume a well-mixed homogeneous environment in which droplets and aerosols are equally spread in space~\citep{Bazant2020,bazant2021guideline}. 
Others examined specific cases such as the efficiency of different masks in blocking the spread of droplets~\citep{mittal2020mathematical, verma2020visualizing} and the dispersion of exhaled droplets during outdoor and indoor activities in different flow environments~\citep{chong2021extended,Ng2021, zhang2021disease,liu2021simulation,nazari2021jet}.

It is now obvious that different flow conditions may spread relatively large exhaled droplets farther than 2 meters.
In such cases, the effectiveness of social distancing, barriers, and ventilation, intended for suppressing the spread of viral diseases might vary significantly, as the dispersion of droplets relies on the specific flow condition. The conditions, in turn, depend on multiple factors (ventilation rates, air conditions, moving objects and vehicles, flow barriers, etc.).
On the other hand, flow conditions generated by expiratory events may lead to the persistence of virus-loaded droplets in the flow, thus producing regions in which the risk of infection via the "airborne droplet" route increases dramatically \citep{Wang2021a}.
This interrelated dependency between the settling distance, virus exposure, and the flow conditions may lead to inconclusive and ineffective recommendations by health regulators \citep{smith2020aerosol}. Notably, \cite{Wang2021a} computed the density of virus copies moving past a defined control area, rather than accounting for the number of droplets.
This approach produced useful exposure maps and highlighted the risk in short-range exposure,  underestimated by the current guidelines.

Therefore, a fundamental understanding of the underlying mechanisms of such multiphase interactions is important in order to accurately model and control the spread of exhaled droplets and aerosols.
\cite{Cummins2020}, for example, studied the dynamics of spherical droplets in the presence of a source-sink pair potential flow field. A mathematical analysis was presented, offering a theoretical model for droplet dispersion using an analytically described carrier flow field.
The study demonstrated the interactions between particles, gravity, and the carrier potential flow, and revealed an intermediate range of droplet diameters for which their horizontal distances are minimized. 

As previously discussed, preventive measures such as ventilation systems and flow barriers, or even indoor movements of objects, are inducing vortical flow structures that affect the spread of micron-sized particles and droplets.
\cite{renzi2020} investigated the role of vortex rings in the settling of exhaled respiratory droplets. Using a theoretical approach and assuming Hill's equation as the carrier flow field, they found that vortex flows may delay the settling time of a particle suspended within an exhaled jet and thus enhance its displacement.

\cite{Monroe2021} studied the vortical flow structures created by pulsating coughing events and the influence of secondary and tertiary expulsions on particle dispersion and penetration. Without accounting for the thermodynamic effects, the overall dispersion of a particle cloud was found to be suppressed by increased pulsatility, while the entrainment of particles ejected during the secondary or tertiary expulsion is enhanced.

 Moreover, \cite{chong2021extended}, \cite{Ng2021}, \cite{Rosti2021}, and \cite{Wang2021} investigated the role of humid, turbulent  puff ejected during a typical respiratory event on the extension of respiratory droplets' lifetime. The studies reported that small droplets might be engulfed inside the turbulent puffs, leading to a significant increase in their lifetimes. This result implies that smaller droplets might be transported much further than expected during the respiratory events. 

 Liu et al. \citep{Liu2021,Liu2021a} explored the fluid dynamics of a turbulent, ejected laden puff resulting from a cough or sneezing. A large-eddy simulation was used to analyze the fluid phase, while the point-particle Euler–Lagrange approach was chosen for tracking the respiratory droplets. Notably, they observed the detachment of a small vortex ring-like structure that advanced relatively large distances and was observed carrying along part of the suspended droplets.

Vortical flows also play a significant role in medical sprays, engineering, and environmental applications. The stability of flows is may be affected by large recirculation zones in reacting and non-reacting flows~\citep{dagan2016, taamallah2019helical, chakroun2019flamelet, DAGAN2019368}. Our recent studies reveal the interactions of such unstable flows and evaporating reacting sprays~\citep{dagan2015dynamics}.
Mathematical descriptions of such interactions were derived to solve the coupled vortex-particle system~\citep{daganILASS2016, daganILASS2017, daganSimilarityFlames2018}, and may also assist in understanding and interpreting results from numerical simulations~\citep{dagan2017particle}. 

Other vortical flow forms of particular interest are dipolar vortices that are ubiquitous in both closed and open environments.
Almost any relative air movement about an object is expected to create such vortical flow structures in its wake, depending on its geometry and flow conditions.
Notably, once created, these flow structures self-propel through space and sustain for relatively long periods.

In our recent study \citep{Dagan2021}, we used the two-dimensional Lamb-Chaplygin dipole solution to examine the effects of vortex dipoles on the dispersion and settling distance of non-evaporating particles. 
In contrast to previous works in this field, we assumed that the cloud of droplets or aerosols is initially at rest and not advected by an exhaled respiratory jet. 
This configuration may serve as a general model for respiratory droplets suspended in the ambient air, for example, after an exhalation event has occurred, and might be utilized as a tool for understanding the possible translation of a cloud of droplets or aerosols even after the exhalation event. 
Moreover, it may allow us to isolate phenomena that are not directly linked to vortical flows and thus explore the fundamental physical principles underlying the entrainment of the particles and their enhanced settling distance.~
However, the previous study \citep{Dagan2021} did not consider the heat and mass transfer between the droplet and the ambient humid air.

Previous works \citep{Wells1934,Xie2007,Wang2020,DeOliveira2021} have demonstrated the significant influence of droplet evaporation rates on the airborne droplet lifetime and settling distances. 
The evaporation of droplets is a fundamental process in aerosol science and is essential in many atmospheric and environmental studies. The interdependence between evaporation and settling of airborne droplets was studied extensively by Kulmala~\citep{Kulmala1989,Kukkonen1989,Kulmala1991} who offered an analytical model for the evaluation of continuum heat and mass flux at the droplet surface. The model includes a description of forced convective mass and heat transfer due to the droplet translation and is therefore well suited for the evaluation of exhaled respiratory droplet evaporation in the vicinity of vortical carrier flows.

\cite{Xie2007} revised the classical Wells evaporation-settling curve and reexamined the droplet evaporation. They considered the effect of various air properties, such as ambient temperature, local velocity, and relative humidity, as well as droplet properties, including initial temperature and salts concentrations. Consequently, these properties were established as important for the accurate analysis of exhaled respiratory droplets~\citep{DeOliveira2021}.

The objective of the present study is to theoretically investigate the effects of dipolar vortex flow structures on the dynamics and deposition of evaporating droplets and solid particles initially suspended in the air.

A mathematical formulation for the vortex carrier flow is presented in Sec.\ref{sec:flow}. In Sec.\ref{sec:level4}, we describe the Lagrangian model, spatially tracking the droplet's velocity, mass, and temperature.
Validation of the numerical procedure is introduced in Sec.\ref{sec:v&s}, and the results are presented in Sec.\ref{sec:results}, for a cloud of exhaled droplets near a vortical flow structure, where the trajectories of the Lagrangian droplets and solid particles are analyzed and studied.
Finally, Sec.\ref{sec:conc} includes concluding remarks and a brief outlook.

% SECTION 2
\section{\label{sec:math} Mathematical model}

A cloud of discrete water droplets initially suspended in undisturbed air is considered.
Due to the movement of objects or flow indoors near the droplets, a wake flow structure might encounter and disperse the droplets.
For low-to-moderate Reynolds numbers and short characteristic times, we may approximate the wake flow as an ideal dipolar vortex. Such vortices may be modeled by the Lamb-Chaplygin dipole~\citep{Swaters1988,Meleshko1994}. 
The effect of this vortical flow structure on the dispersion of the exhaled respiratory droplet is realized by considering the coupled aerodynamic and thermodynamic properties of individual Lagrangian particles, carried by an analytically described flow field. The equations for the carrier flow and Lagrangian particles are written as follows.
\subsection{\label{sec:flow} Carrier flow-field}
\begin{figure}
  \centerline{\includegraphics[width=0.65\linewidth]{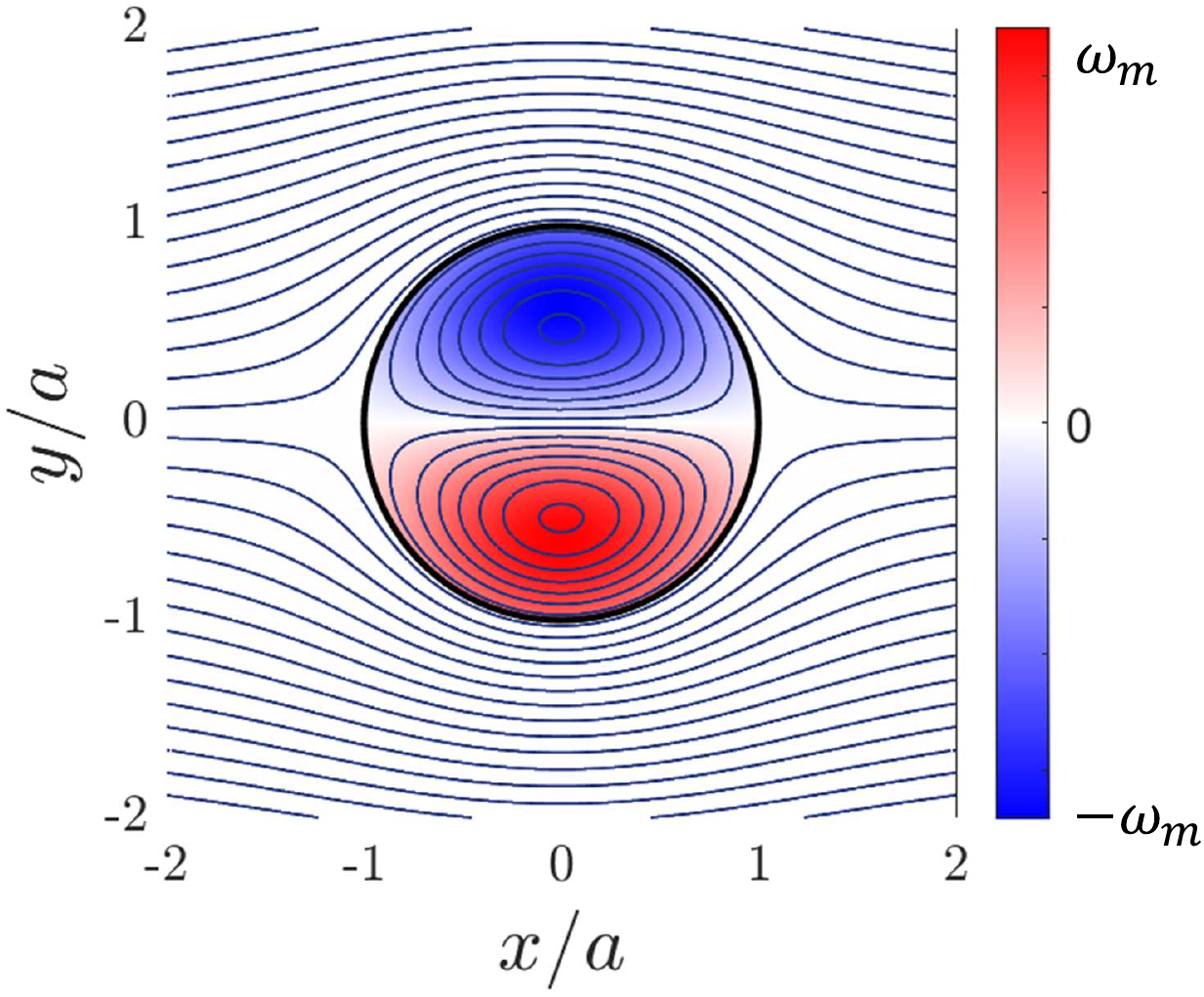}}
  \caption{Lamb-Chaplygin vortex dipole streamlines and qualitative representation of the vorticity field $\omega(\bar{x},\bar{y})$, illustrated at the normalized frame of reference fixed to the dipole center. The bold black line represents the transition between the inner, vortical flow and the outer flow.}
  \label{fig:strm}
\end{figure}

We seek to model self-propelling, low velocity, two-dimensional bipolar vortex flows in an indoor environment.
The viscous decay of such vortex flows was studied experimentally by \cite{Flor1995}, who analyzed the vortex decay due to the diffusion of vorticity and entrainment of ambient irrotational fluid surroundings. 
Considering a circular dipole structure of a constant radius $a$ moving at a steady translation velocity of $U_0$ in a quiescent fluid with dynamic viscosity $\nu_{f}$, the decay timescale was found as $t_v \approx a^2 / \nu_f$. 
Thus, as long as $t \ll t_v$, we deem the carrier flow as laminar and ideal and consider the translation velocity to be constant.
A solution for this idealized flow was derived by Lamb \citep{Swaters1988} and Chaplygin \citep{Meleshko1994}, for radial coordinates $(r,\theta)$ positioned at the center of the vortex core in a frame of reference moving with the vortex:  
\begin{equation}
    \psi(r,\theta)=\frac{-2aU_0\sin(\theta)}{bJ'_1(b)}J_1 \left (\frac{br}{a} \right),\;\;\;\;r\leq a,
\end{equation}
\begin{equation}
    \psi(r,\theta)=U_0\sin(\theta)r \left (1-\frac{a^2}{r^2} \right),\;\;\;\;r>a,
\end{equation}
where $J_1,J'_1$ and $b$ are the first-order Bessel function of the first kind, its derivative, and its smallest non-trivial root $b=3.8317$, respectively.
Using the stream function, the radial and azimuthal velocities are given as:
\begin{equation}
    \begin{split}
        &u_r=-2a U_0 \cos(\theta) \frac{J_1(\frac{br}{a})}{brJ'_1(b)}\,;\\ &u_{\theta}=-2U_0 \sin(\theta) \frac{J'_1(\frac{br}{a})}{J'_1(b)},\;\;\;\;r\leq a,    
    \end{split}
    \label{eq:flowvel1}
\end{equation}
\begin{equation}
    \begin{split}
        &u_r=U_0 \cos(\theta) \left(1-\frac{a^2}{r^2} \right )\,;\\
        &u_{\theta}=-U_0 \sin(\theta)\left(1+\frac{a^2}{r^2} \right),\;\;\;\;r>a.
    \end{split}
    \label{eq:flowvel2}
\end{equation}
The vorticity field varies inside the core, as the points of maximal vorticity are located at $\delta=\pm 0.48a$, as illustrated in Fig.\ref{fig:strm}. The maximal intensity might be written in terms of the translation velocity and radius as:
\begin{equation}
    \omega_m=11.09\frac{U_0}{a}.
\end{equation}
A streamline representation of the Lamb-Chaplygin vortex is illustrated in Fig.\ref{fig:strm}, in the normalized vortex frame of reference $(\bar{x},\bar{y})$, alongside the vorticity field intensity distribution.
We use the above solution as the background flow field, dispersing the droplets exhaled in the vicinity of the dipole.

\subsection{\label{sec:level4} Single droplet Lagrangian equations}

For the discrete phase, we consider a dilute, micron-sized droplet cloud initially at rest. As such, we may assume the unsteady motion of the droplets does not affect the carrier flow field.
Furthermore, we shall discount any potential interactions between the droplets, including collision and coalescence, and assume that the presence of nearby droplets does not affect the properties of the air surrounding each droplet.
Under these assumptions, the Lagrangian conservation equations for single droplets are introduced and shall be employed for the computation of the droplet spatial location $\bar{x}_d$, velocity $\bar{u}_d$, diameter $d$, and temperature $T_d$. 

\subsubsection{\label{sec:eq_mom} Equations of Motion}
The general form of the equations of motion for small particles in nonuniform, unsteady flows were derived by \cite{Maxey1983}, taking into consideration gravity, drag, virtual mass, and the Basset "history" force.
This study concerns a small water droplet motion in the air, where the particle-medium density ratio is large, and the particle characteristic length is much smaller compared to the flow integral length.~ Hence, we may postulate that the magnitude of the forces due to undisturbed flow, virtual mass, Faxen's drag correction, as well as lift forces and particle history terms are of couple order smaller than the magnitude of the drag and gravity forces. We may reduce the general form of the equations to the following:
\begin{equation}
    \frac{d\bf x_d}{dt}=\bm{u_d},
    \label{eq:location}
\end{equation}
\begin{equation}
    \frac{d\bm{u_d}}{dt}=18f\frac{\nu_{f}\rho_{f}}{\rho_{d}d^2}(\bm{u_f}-\bm{u_d})+\bm{g},
    \label{eq:moment}
\end{equation}
where $\bm{x_d}$ and $\bm{u_d}$ are the droplet location and velocity vectors in the vortex frame of reference, $\bm{g}$ is the gravitational acceleration, and $\bm{u_f}$ is the background flow local velocity, given by Eq. \ref{eq:flowvel1} and \ref{eq:flowvel2}. 
Here, $\rho_{f}$ is the fluid (air) density, $\rho_{d}$ is the droplet density and $f$ is the drag factor- the ratio of the drag coefficient to Stokes drag coefficient.
$f$ was found empirically by several authors, as a function of the droplet's relative Reynolds number- $Re_d=d|\bm{u_f}-\bm{u_d}|/\nu_f$. For $Re_d$ up to 800, suiting our wake flow regime, the drag factor was found by Schiller and Naumann \citep{Crowe2011} as:
\begin{equation}
    f=1+0.15Re_{d}^{0.687},
\end{equation}
Although the carrier flow field is rotational, our Lagrangian model does not account for angular momentum conservation, hence neglecting the droplet spin and its influence over the droplet's dynamics. 

\subsubsection{\label{sec:eq_mass} Transport Equations}
The diameter and temperature of the Lagrangian droplet are governed by heat and mass transfer processes, as it might undergo a phase change such as evaporation, condensation, or even freezing. 
As we aim to model exhaled respiratory droplets scattering in an indoor environment, where we can safely assume freezing is not a relevant phase-change process.

~ Kulmala \citep{Kulmala1989,Kulmala1991} have formulated the mass transfer at the droplet-air interface for a quasi-stationary case, assuming the droplet-air interface is saturated, and a zeroth-order mass fraction profiles in the continuum phase.
Although diffusivity was considered as the sole transfer mechanism, an expansion of the solution for ventilated droplets might be carried out using the Sherwood dimensionless number $Sh$, defined as the ratio of total mass transfer to the purely diffusive flux.
In terms of droplet diameter, the mass equation for the general case is:
\begin{equation}
    \frac{d}{dt}(d)=4C_TSh\frac{M_{w,v}D_{\infty}p_{\infty}} {RT_{\infty}\rho_d d} \ln\left(\frac{p_{\infty}-p_{v,d}}{p_{\infty}-p_{v,\infty}}\right),
    \label{eq:mass}
\end{equation}

Where $M_{w,v}$ is the molecular weight of the vapor, $R$ is the universal gas constant, $D_{\infty}$ is the ambient binary diffusion coefficient, and $p_{\infty}$ is the ambient pressure.
We take the vapor pressure at the droplet surface $p_{v,d}$ as the vapor pressure corresponding to the droplet temperature $p_{v,d}=p_{sat}(T_d)$, and the ambient vapor pressure $p_{v,\infty}$ as a function of the ambient temperature and relative humidity $p_{v,\infty}=RHp_{sat}(T_{\infty})$. The saturation pressure $p_{sat}(T)$ was calculated using the Goff-Gratch formula \citep{goff1957saturation}.
The term $C_T$ in Eq.\ref{eq:mass} is the diffusion coefficient temperature dependence factor, which is given by
\begin{equation}
    C_T=\frac{T_{\infty}-T_{d}}{T_{\infty}^{\mu-1}} \space \frac{2-\mu}{T_{\infty}^{2-\mu}-T_{d}^{2-\mu}}.
    \label{eq:Ct}
\end{equation}
The term $\mu$ in Eq.\ref{eq:Ct}|~is a substance-specific constant, which ranges between $1.6<\mu<2$ for most substances. 
For water droplets in room conditions, the temperature depression $T_{\infty}-T_{d}$ is relatively small, and thus the deviation in the diffusion coefficient predicted by the factor $C_T$ is smaller than 3\% \cite{Kulmala1989}. Hence, in the present study, we choose to set $C_T=1$ and neglect the variation in the diffusion coefficient.

By accounting for the sensible heat stored in the droplet, heat advection due to mass transport, conduction, and convection at the droplet surface, one may derive the energy conservation equation for a single droplet assuming a uniform temperature distribution:
\begin{equation}
    \frac{dT_d}{dt}= \frac{3h_{fg}}{c_{p,d}}\frac{1}{d}\frac{d(d)}{dt}-Nu \frac{12  k_f}{c_{p,d}\rho_d}\frac{T_d-T_{\infty}}{d^2},
    \label{eq:heat}
\end{equation}
where $h_{fg}$ is the heat of vaporization, $c_{p,d}$ is the droplet heat capacity, and $k_f$ is the fluid heat conductivity, and $Nu$ is the Nusselt dimensionless number.
By applying a dimensional analysis based on the similarity of the mass and heat equation, we may determine  $Sh=Sh(Re_d,Sc),Nu=Nu(Re_d,Pr)$, where $Sc=\frac{\nu_f}{\rho_f D_{\infty}}$ and $Pr=\frac{c_{p,f}\nu_f}{\rho_f k_f}$ are the fluid Schmidt and Prandtl numbers, respectively. Correlations were found experimentally for moderate Reynolds, as \citep{Fuchs1959}:
\begin{equation}
    \begin{split}
        &Sh=1+0.276Re_d^{1/2}Sc^{1/3}\,;\\ &Nu=1+0.276Re_d^{1/2}Pr^{1/3},
    \end{split}
\end{equation}
which will be adopted in the present study.

\subsection{Computational Setup and Model Validation \& Verification} \label{sec:v&s}

Unlike the carrier flow, which is described by the steady analytical Lamb-Chaplygin solution, the coupled ordinary differential equation system is resolved numerically.
We track the droplet's spatial location by solving Eq.\ref{eq:location}, and find its Lagrangian velocity by solving Eq.\ref{eq:moment}. This equation, in turn, is coupled with Eq.\ref{eq:mass} and Eq.\ref{eq:heat}, describing droplet diameter and temperature, correspondingly. The coupled system is solved simultaneously using RK4 method with a variable time step.

The droplet diameter was limited to 10\% of its initial value. By the time the droplet losses more than 99\% of its mass, we assume the droplet would have evaporated completely and thus eliminate it from the simulation. However, \cite{Lieber2021} found that a saliva aerosol may linger in the air for extended periods following the evaporation of the water content. This aerosol generation mechanism is not considered in the present study but might be of significance.

Our numerical scheme was validated and verified with the published results of three distinct study cases.
\cite{Ranz1952} have investigated the evaporation of stationary water droplets ($T_{d,0}=282K, d_0=1.1mm$) in dry air ($T_{\infty}=298K, RH=0\%$), while \cite{Smolik2001} measured the evaporation rate of a suspended droplet ($T_{d,0}=287K,d_0=1.2mm$) in a steady air stream ($T_{\infty}=297K, RH=35\%,U_a=0.2 m/s$). Additionally, \cite{Xie2007} have developed a semi-analytic model and investigated the evaporation and movement of droplets expelled during respiratory activities; in particular, the well-known Wells evaporation–falling curve of droplets.

\begin{figure}
    \centerline{\includegraphics[width=0.65\linewidth]{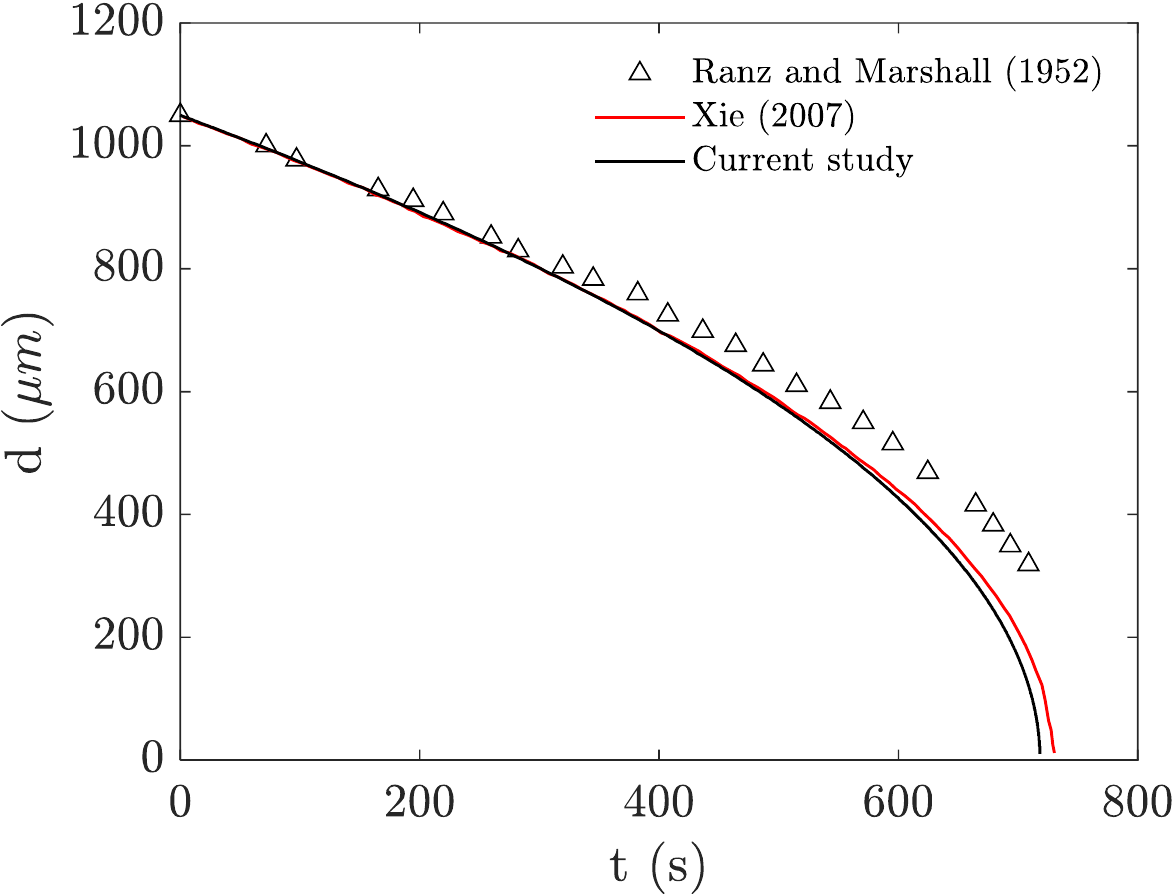}}
    \caption{Comparison of droplet evaporation rates measured by \cite{Ranz1952} (black triangles), the numerical solution of \cite{Xie2007} (solid red line) and current study (solid black line).}
    \label{fig:firstval}
\end{figure}

\begin{figure}
    \centerline{\includegraphics[width=0.65\linewidth]{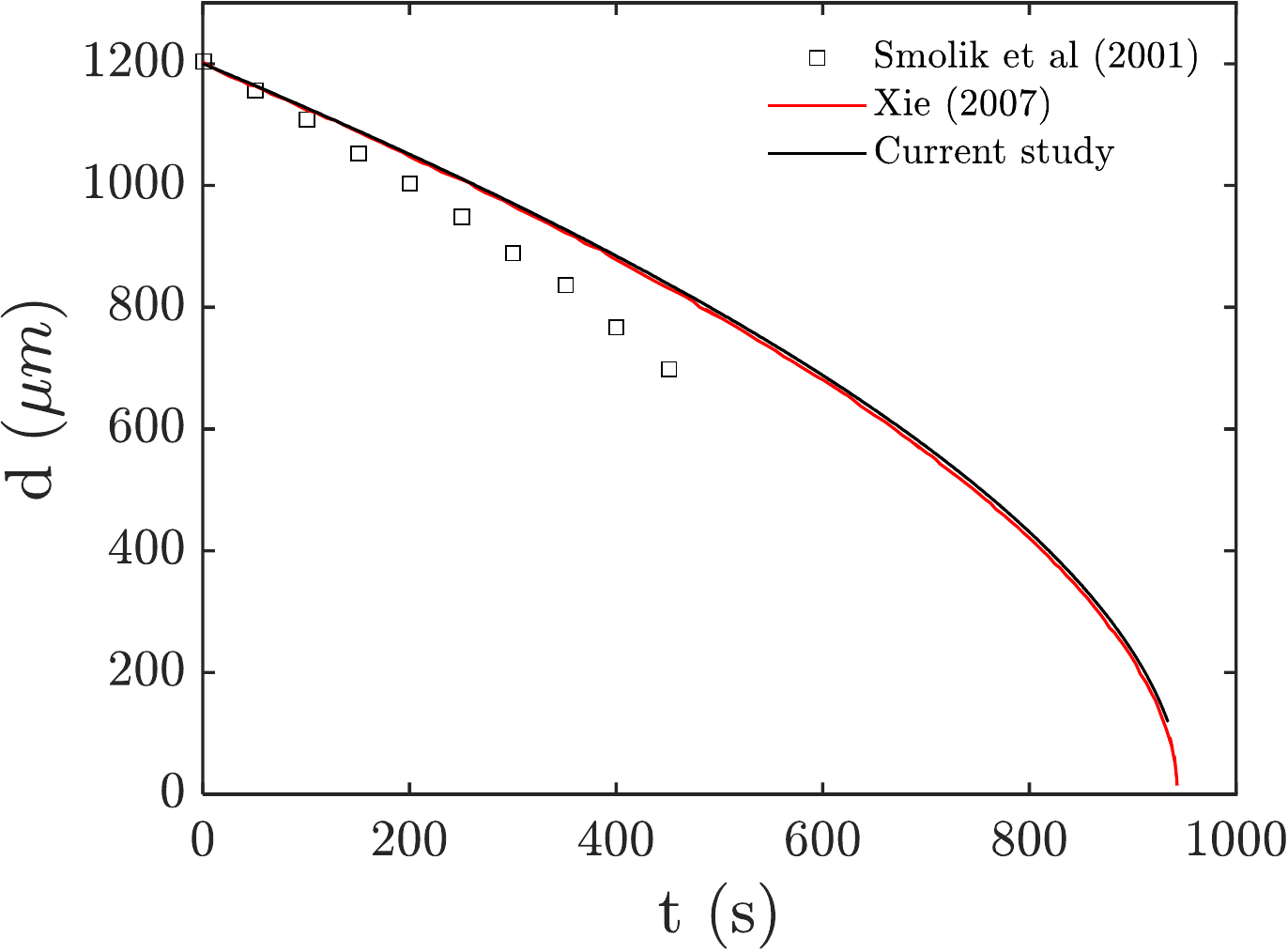}}
    \caption{Comparison of droplet evaporation rates measured by \cite{Smolik2001} (black squares), the numerical solution of \cite{Xie2007} (solid red line) and current study (solid black line).}
    \label{fig:secondval}
\end{figure}

Validation of our suggested model using the results of mentioned experiments, as well as verification of the numerical scheme using the modeling results presented by \cite{Xie2007}, are shown in Fig. \ref{fig:firstval} and Fig.\ref{fig:secondval}.

Both numerical solutions from both studies are almost identical. The demonstrated small discrepancy might stem from the usage of slightly different correlations for the evaluation of dimensionless numbers ($Nu,Sh$), factors ($f$), and thermodynamic properties.
Fig. \ref{fig:secondval} reveals an overestimation of both models for droplets evaporating in a constant air stream after long integration times, while the steady evaporation was predicted more accurately. This suggests the convective transport influence, represented by Nusselt and Sherwood numbers, might be over-predicted by currently used correlations. 

\begin{figure}
    \centerline{\includegraphics[width=0.65\linewidth]{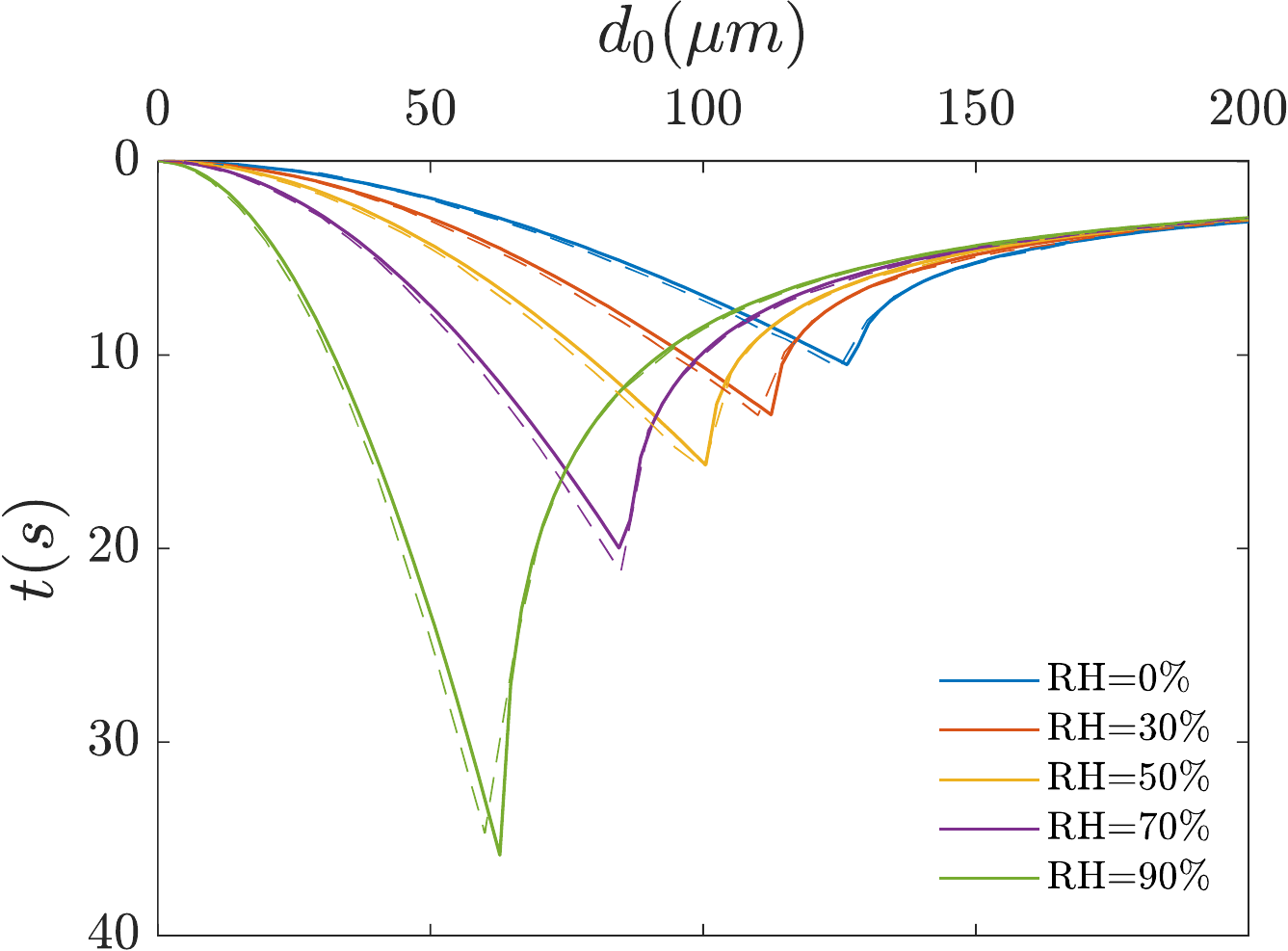}}
    \caption{Airborne lifetime of a $T_{d,0}=306K$ water droplet free-falling in quiescent air ($T_{\infty}=291K$) and various relative humidities, as a function of its initial diameter. Solid lines represent the current study predicaments, and dashed line are modeling results of \cite{Xie2007}.}
    \label{fig:thirdval}
\end{figure}

Wells diagrams for droplets of varying initial diameters ($T_{d,0}=306K$) free-falling from 2 m to the ground in quiescent, moist, air of various relative humidities and ambient temperature of $T_{\infty}=291K$, are plotted in Fig. \ref{fig:thirdval}.
Good agreement between the current study and \cite{Xie2007} numerical results were achieved. The influence of the ambient air properties on the heat and mass transport out of the droplet is showcased here clearly. As expected, the evaporation of droplets is significantly slower in moist air than in dry air, and thus they accumulate more inertia and fall quicker. On the other hand, under such conditions, the airborne lifetime might be enhanced due to slower evaporation, slowing down the drying of the droplet.

\begin{figure}
    \centerline{\includegraphics[width=0.7\linewidth]{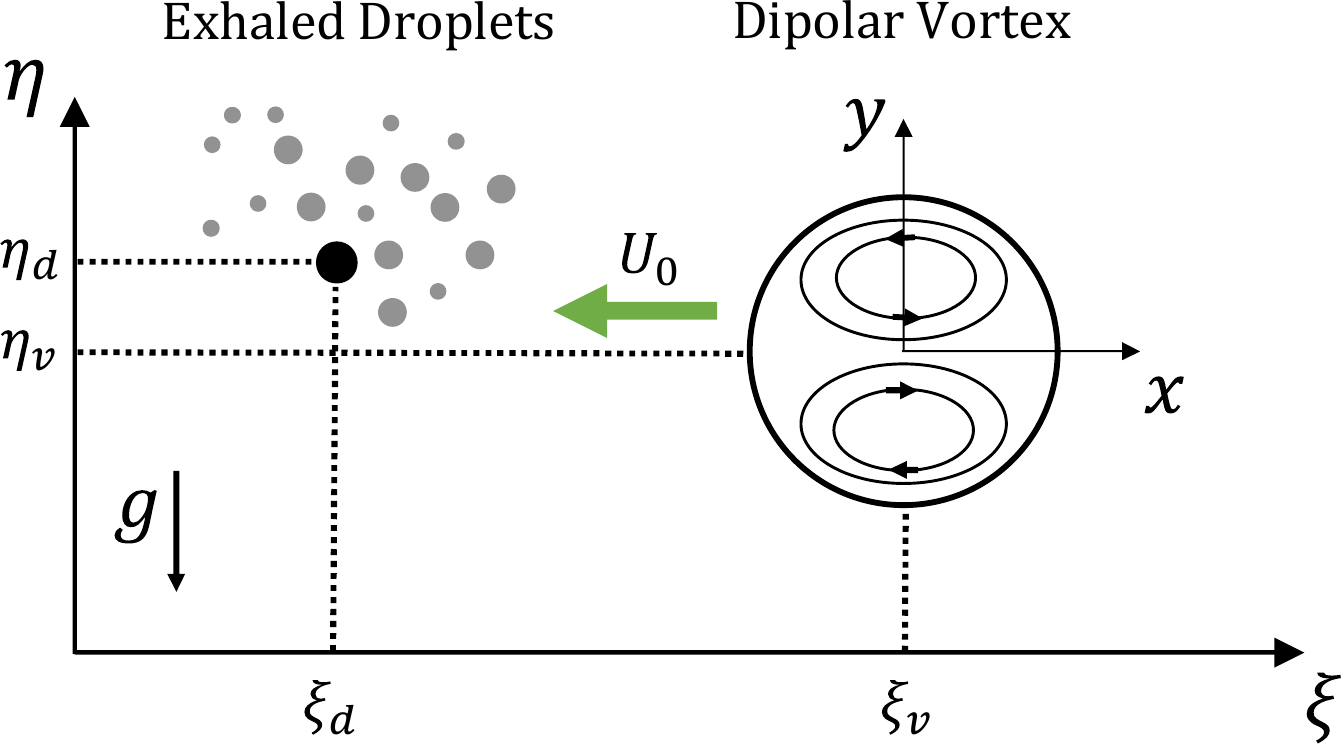}}
    \caption{Flow-particle configuration illustrated with respect to the stationary frame of reference $(\xi, \eta)$. Droplets, initially placed at $(\xi_d, \eta_d)$, free fall due to gravity $\bar{g}$. A vortex dipole flow-field, initially placed at $(\xi_v, \eta_v)$, is self-propelling towards the free-falling droplets at a velocity $U_0$.}
    \label{fig:conf}
\end{figure}

We will examine a self-propelling vortex dipole in an ambient air moving parallel to the ground towards a cloud of droplets, as illustrated in Fig.\ref{fig:conf}.
We consider the vortex is moving at the height of $\Delta \eta _{v}=2m$ from the ground, located initially at $\xi_{d}$, and analyze its influence over free-falling trajectories and the settling times of a single droplet, located at $(\xi  _ {d},\eta _ {d})$ relative to the vortex core.

Modeling indoor dispersion of small respiratory droplets, we shall investigate droplets of initial diameters between $1 \mu m $ and $200 \mu m $ \citep{smith2020aerosol}.
Moreover, we may take the ambient air temperature to be $T_{\infty}=298K$, the relative humidity to be $RH=50\%$, and the droplet initial temperature to be $T_{d,0}=308K$.  The thermodynamic conditions are fixed for all the presented results.  A vortex size of $a=0.1m$ is considered, representing a typical wake structure due to the motion of objects in a room, e.g., a hand movement. Following the same logic, the propagation velocity was assumed to be roughly $U_0\approx 0.1-0.5 m/s$ \citep{Luo2018}, and thus we studied dipoles with maximal vorticity in the range of up to $\omega_m=50 \hspace{0.1cm} 1/s$.

It should be noted that the vortex translational velocities realized here are low relative to the maximum velocity of exhaled respiratory events such as sneezing and coughing. However, by studying the settling distances of a cloud of particles that is initially at rest, we may theoretically simulate the case of droplet fallout following a violent respiratory event.

The simulation domain was chosen as $\xi\in[-8a,a]$ and $\eta\in[-20a,3a]$ in the stationary reference and relative to the dipole center. Initial dynamic conditions for the droplets are zero velocity in both directions; as we take to be sufficiently distant, so its effect on the droplets is negligible.

\begin{figure*}  
    \centering
    \centerline{\includegraphics[width=1\textwidth]{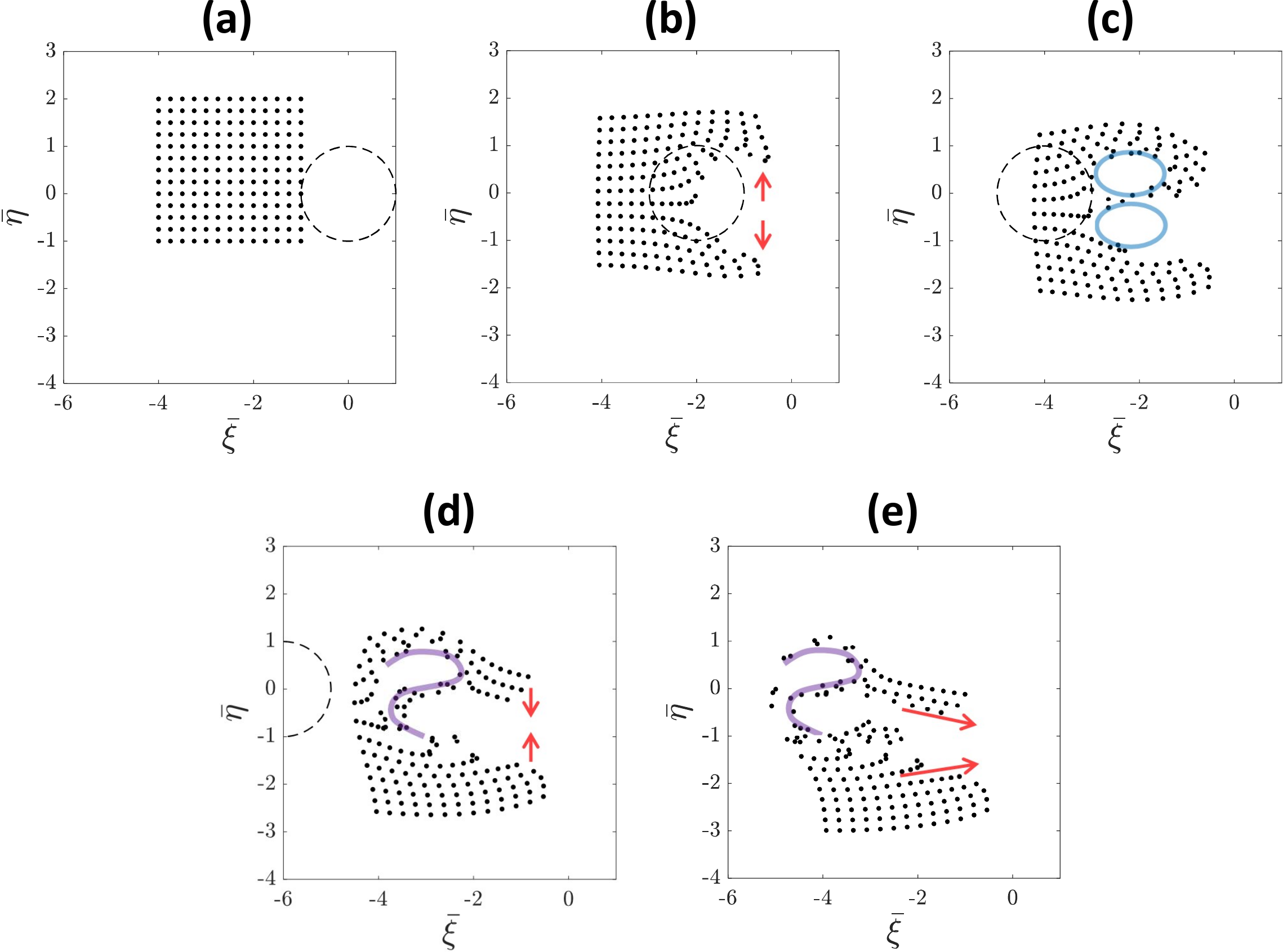}}
    \caption{Instantaneous snapshots of the dispersion caused by a vortex of maximal intensity $\omega _{m} =25\hspace{0.1cm} 1/s$ passing through a dilute cloud of $d_0=65 \mu m$ droplets. The dashed line represents the vortex core, moving from right to left. Subplots (a-e) illustrate the cloud dispersion at a different time: $t=0,~2a/U_0,~4a/U_0,~6a/U_0,~8a/U_0$, Both axes are normalized with respect to the vortex radius $a$.}
     \label{fig:matrix}
\end{figure*}

% SECTION 3
\section{Results} \label{sec:results} 
The dispersion of a droplet cloud by a passing vortex structure is illustrated in Fig.\ref{fig:matrix}, in a stationary frame of reference $(\xi,\eta)$. Both axes are normalized by the vortex radius $a$. The vortex was chosen such that its maximal intensity is $\omega_m=25 \hspace{0.1cm} 1/s$, representing the characteristic wake flow structure. 
Initially (Fig.\ref{fig:matrix}a), the cloud consists of discrete, $d_0=65\mu m$ water droplets, represented as a two-dimensional array of droplets scattered in the region $-4<\xi<-1$, $-1<\eta<2$ outside the vortex core.
Figures \ref{fig:matrix}b-e illustrates the evolution of the droplet's spatial location at evenly spaced time instances, each corresponding to the vortex propelling one diameter parallel to the ground.

Fig.\ref{fig:matrix}b demonstrates the initial dispersion of the cloud.
As the vortex moves through the array, it pushes most of the droplets outwards and around it (red arrows), creating void regions behind the vortex.
Yet, some droplets penetrated the dipole core, as the carrier flow could not advect them around it.
At this point, the symmetry of the cloud is mostly conserved, as the droplets are pushed rather equally in both directions.
When the vortex keeps progressing, the symmetry of the cloud seems to break, as observed in \ref{fig:matrix}c.
A difference between the upper and lower parts of the cloud is noticeable, as gravity alters the scattering pattern made by the symmetric carrier flow field.
Furthermore, dilute regions within the cloud are appearing (blue circles), corresponding to the high-vorticity regions inside the dipole (Fig.\ref{fig:strm}).
The droplets are arranged around these dilute regions, such that the dipole zero-vorticity centerline is the only region that remained occupied by droplets.
One may notice that distinct S-shaped traces of droplets are formed, highlighted in Fig.\ref{fig:matrix}d (purple line), as a result of the droplet's movement around the high-vorticity regions. These traces in the cloud are maintained even after the vortex passes the vicinity of the cloud, as illustrated in Fig.\ref{fig:matrix}e.
Opposite to the initial stage, when the vortex exits the array, it causes a suction at the back as the droplets are pulled towards the vortex centerline. This behavior is illuminated using red arrows in \ref{fig:matrix}d-e.

As discussed above, the vortex disperses the initially ordered array of droplets, resulting in the highly asymmetric cloud presented in Fig.\ref{fig:matrix}e. As some droplets are advected by the vortex significantly, others are nearly free-falling and settle vertically. Hence, we shall investigate and analyze the influence of the vortex characteristics on single droplets' trajectories and settling distances. 

\begin{figure*}
    \centering
    \centerline{\includegraphics[width=1\textwidth]{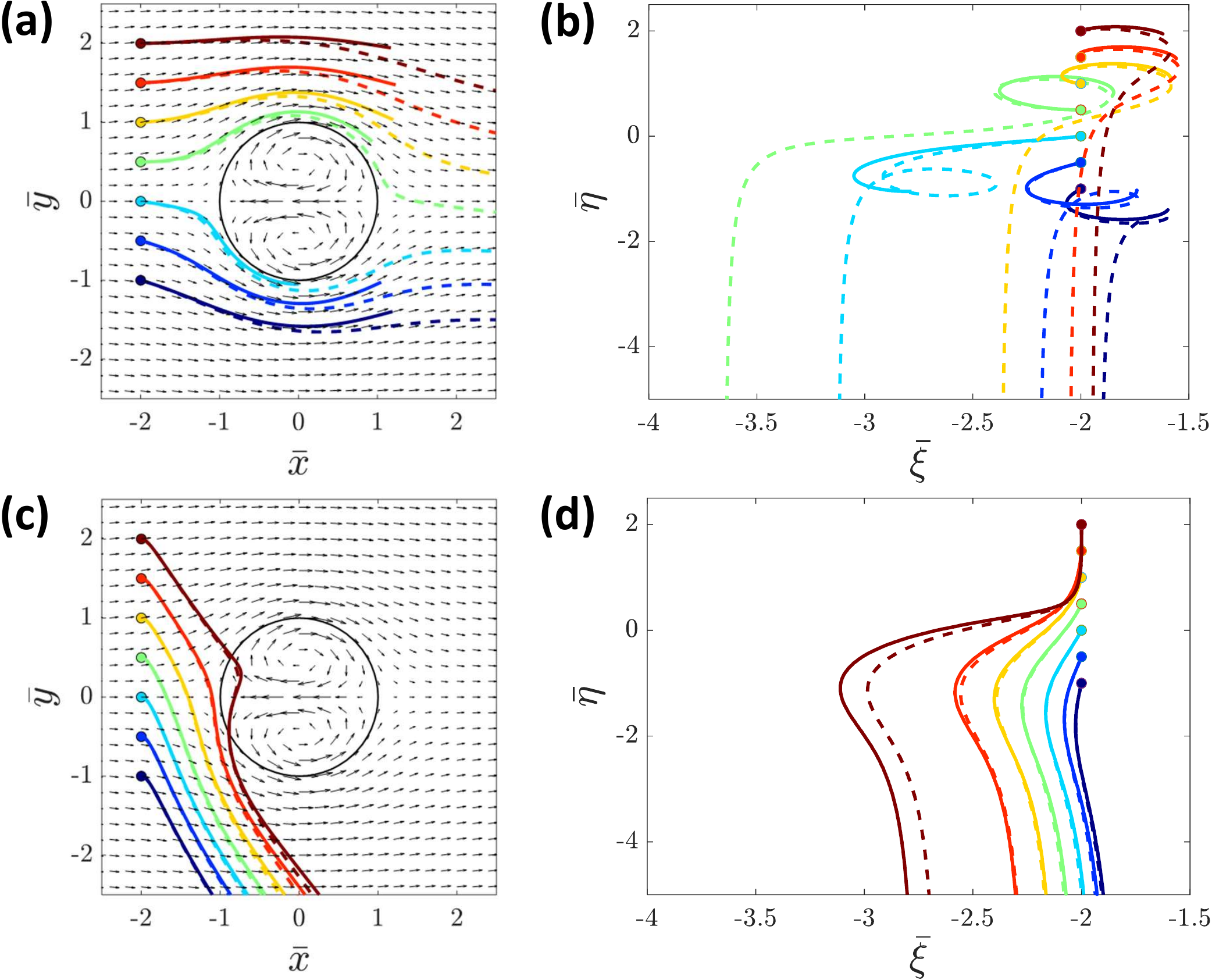}}
    \caption{Comparison of the droplet trajectories (solid lines) and particle trajectories (dashed lines) in the vortex frame of reference having two different initial diameters: $d_0=30\mu m$, and $d_0=130\mu m$ (a,c). The velocity field of the $\omega _{m} =25 \hspace{0.1cm} 1/s$ vortex is presented in the vortex frame of reference, along with the diameter of the vortex core (black line). Colored bullets denote the starting point from which droplets are released, as each color represents a different initial location. (b,d) presents the corresponding trajectories in the stationary frame of reference. The carrier flow is directed leftwards.}
    \label{fig:path1}
\end{figure*}

\begin{figure*}
     \centering
     \centerline{\includegraphics[width=1\textwidth]{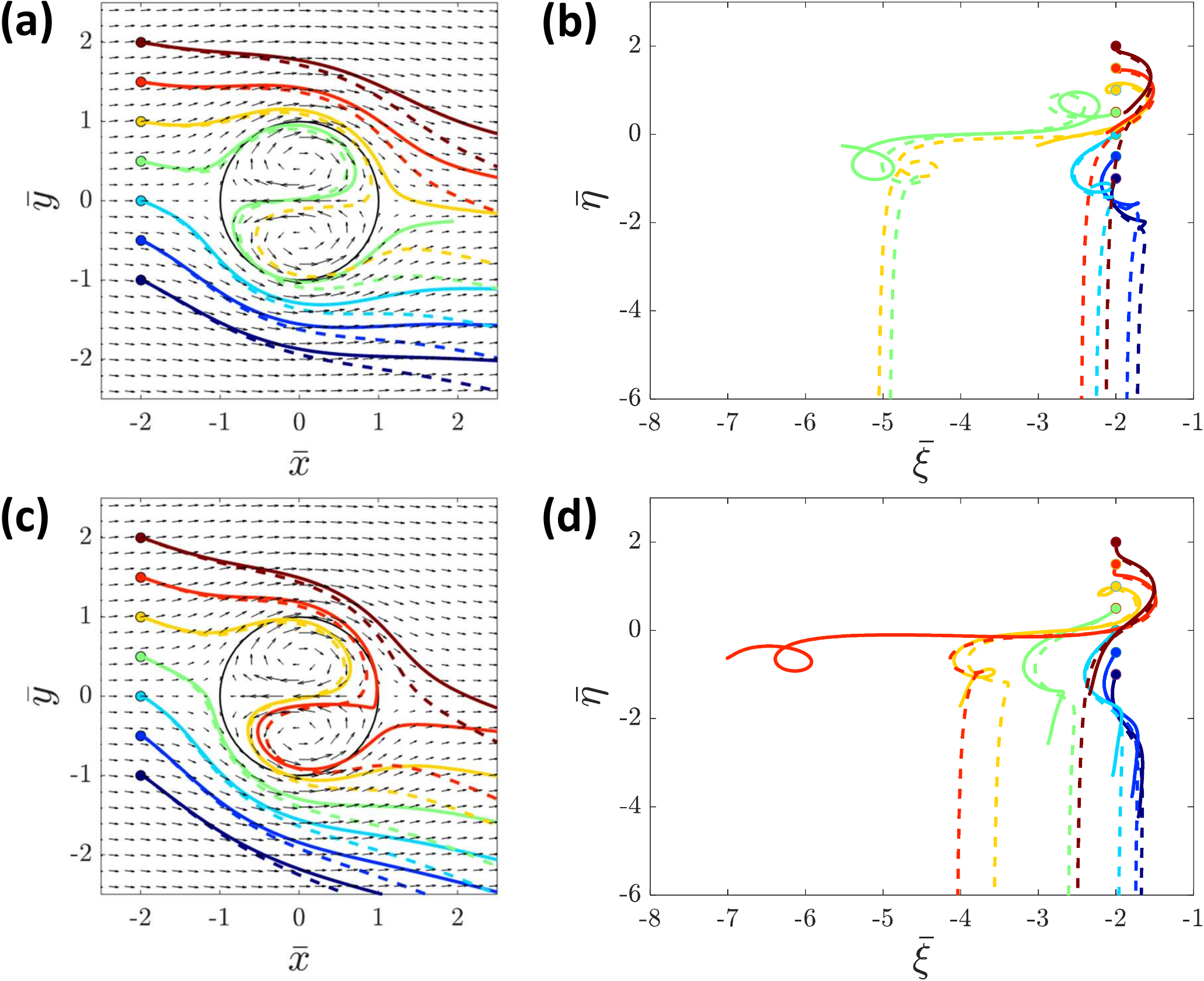}}
        \caption{Comparison of the droplet trajectories (solid lines) and particle trajectories (dashed lines) in the vortex frame of reference having two different initial diameters: $d_0=50\mu m$, and $d_0=65\mu m$ (a,c). The velocity field of the $\omega _{m} =25 \hspace{0.1cm} 1/s$ vortex is presented in the vortex frame of reference, along with the diameter of the vortex core (black line). Colored bullets denote the starting point from which droplets are released, as each color represents a different initial location. (b,d) presents the corresponding trajectories  in the stationary frame of reference. The carrier flow is directed leftwards.}
        \label{fig:path2}
\end{figure*}

A key feature of our model is the coupling of the transport phenomena, the carrier flow, and droplet dynamics. To test the influence of evaporation on the droplet settling distance, we examine the trajectories of droplets of various initial diameters and compare them to the paths of solid particles of the same diameter that do not undergo any phase change, hence maintaining a constant diameter.
Figures \ref{fig:path1} and \ref{fig:path2} show the trajectories of droplets and particles having various initial diameters, released at $x=-2a$ and seven evenly spaced vertical positions $y$, relative to the center of the abovementioned dipole structure.
In the vortex frame of reference (subplots a and c), the carrier flow velocity field is illustrated by a vector plot (black arrows). Droplets are seen as traveling towards the vortex from left to right, carried by the flow field. Trajectories of both solid, non-evaporating particles (dashed lines) and evaporating droplets (full lines) are compared. The corresponding paths are shown in a stationary frame of reference in subplots (b) and (d).

Fig.\ref{fig:path1}a reveals that the lightest particles and droplets ($d_0=30\mu m$) closely follow the dipole outer streamlines. This is expected, as small particles have not accumulated enough inertia to cross the carrier flow streamlines.
Although gravity has not affected the observed paths significantly, it did induce a slight asymmetry between the trajectories of particles released above and below the centerline.

The key influence of the evaporation process, enhanced by the carrier flow, is further demonstrated in Fig.\ref{fig:path1}b. The trajectories of the droplets are similar to the path lines of the respective particles but are much shorter, as the droplets evaporate quickly and disappear completely while still in the vicinity of the vortex core. The distinction between the droplets and particles paths is clear, as the short lifetime of the droplets leads to mid-air evaporation, rather than settling on the ground as the particles. 
One may notice the clockwise loop performed by the four uppermost particles as they are released, while the bottom particles exhibit a counter-rotating motion.
This motion pushes the particles near the centerline further downstream (green and cyan trajectories), greatly extending the particle settling distance.
However, the magnitude of the entrainment decays significantly as the particles are placed farther away from the center and even turns negative for the particles at the outer fringes, i.e., they are pushed backward, opposite to the vortex translation (brown and dark blue trajectories).

Per contra, Fig.\ref{fig:path1}c-d demonstrate the behavior of the heaviest droplets ($d_0=130\mu m$).
As anticipated, both particles and droplets are less responsive to the carrier flow field due to their increased gravitational inertia, and as a result, they settle down quickly and disperse to shorter distances. 
The convergence of trajectories between most of the solid particles and the evaporating droplet trajectories is clear. Here, the droplets' initial mass is large enough such that their evaporation does not alter the mass significantly prior to settling, and thus the observed motion does not differ substantially from the motion of the particles.
However, the top droplet (brown) deviated from the general trend exhibited by the rest of the droplets, having settled farther than the corresponding particle. The top droplet has a longer settling time, and hence both the particle and the droplet have reached the dipolar core with higher inertia. Therefore, both were able to penetrate the vortex and entrained by it downstream. In the process, the droplet had evaporated and lost mass compared to the particle, leading to it being affected by the vortex to a greater extent and ejecting farther downstream.

While Fig.\ref{fig:path1} presented the trajectories of the smallest and largest droplets, Fig.\ref{fig:path2} presents the dynamics of intermediate diameter particles and droplets of $d_0=50\mu m$ (a-b) and $d_0=65\mu m$ (c-d).
At these conditions, the complex and nontrivial interaction between the droplets and the vortex is observable, revealing unanticipated phenomena.
Fig.\ref{fig:path2}a reveals that particles placed at $\bar{y}_0=0.5$ and $\bar{y}_0=1$ manage to penetrate the vortex core. On the one hand, the particles have enough inertia to cross from the outer region of the flow to the vortex core.
On the other hand, the particles are sufficiently light to be caught between the two counter-rotating vortices and swirl inside the vortex core. This, in turn, increases the particle's residence time within the vortex, displacing the particles significantly while translating along with it, as evident in Fig.\ref{fig:path2}b.
Interestingly, the trajectory of this particle resembles the S-shaped traces developed in the droplet cloud (Fig.\ref{fig:matrix}). 

Evaporation has a "nonlinear" influence over the droplet settling distance in comparison with the parallel particle, as some droplets settle farther, while others exhibit the opposite trend.
First, we shall analyze the trajectory of the droplet initially placed at $\bar{y}_0=1$ (yellow line). Clearly, there is a large difference in displacement compared to the matching particle.
The difference is due to the fact that the droplet loses inertia while evaporating, which in turn may prevent the droplet from penetrating the core, reducing the droplet-vortex interaction, and leading to lesser dispersion.
Notably, the opposite trend is demonstrated in Fig.\ref{fig:path2}c-d. Here, one may notice that the droplet initially placed at $\bar{y}_0=1.5$ (red line) was displaced more than twice as far as the particle placed at the same location. Both the droplet and the particle were caught between the vortices and escaped the core at the same location. 
The process leading to this phenomenon can be illustrated as follows: since the droplet is losing mass, it penetrates the core slightly to the right compared to the particle and reaches the low-velocity region behind the dipole.
When suspended there, the droplet is carried downstream along with the vortex as the relative velocity vanishes and the droplet is entrapped until gravity drives it outside the core. Moreover, the decrease of the relative velocity leads to a lower evaporation rate as the convective heat and mass transfer rates are suppressed, thus delaying the extinction of the droplet and allowing it to travel farther.
The stagnant region is amplifying the disparity between the trajectories even after the droplet has escaped the core. As the droplet evaporates, it is less affected by gravity, thus able to keep circulating the lower vortex. So, unlike the particle, it is ejected upwards, en route the low-velocity region again, and thus pushed even farther downstream.

\begin{figure}
  \centerline{\includegraphics[width=0.65\linewidth]{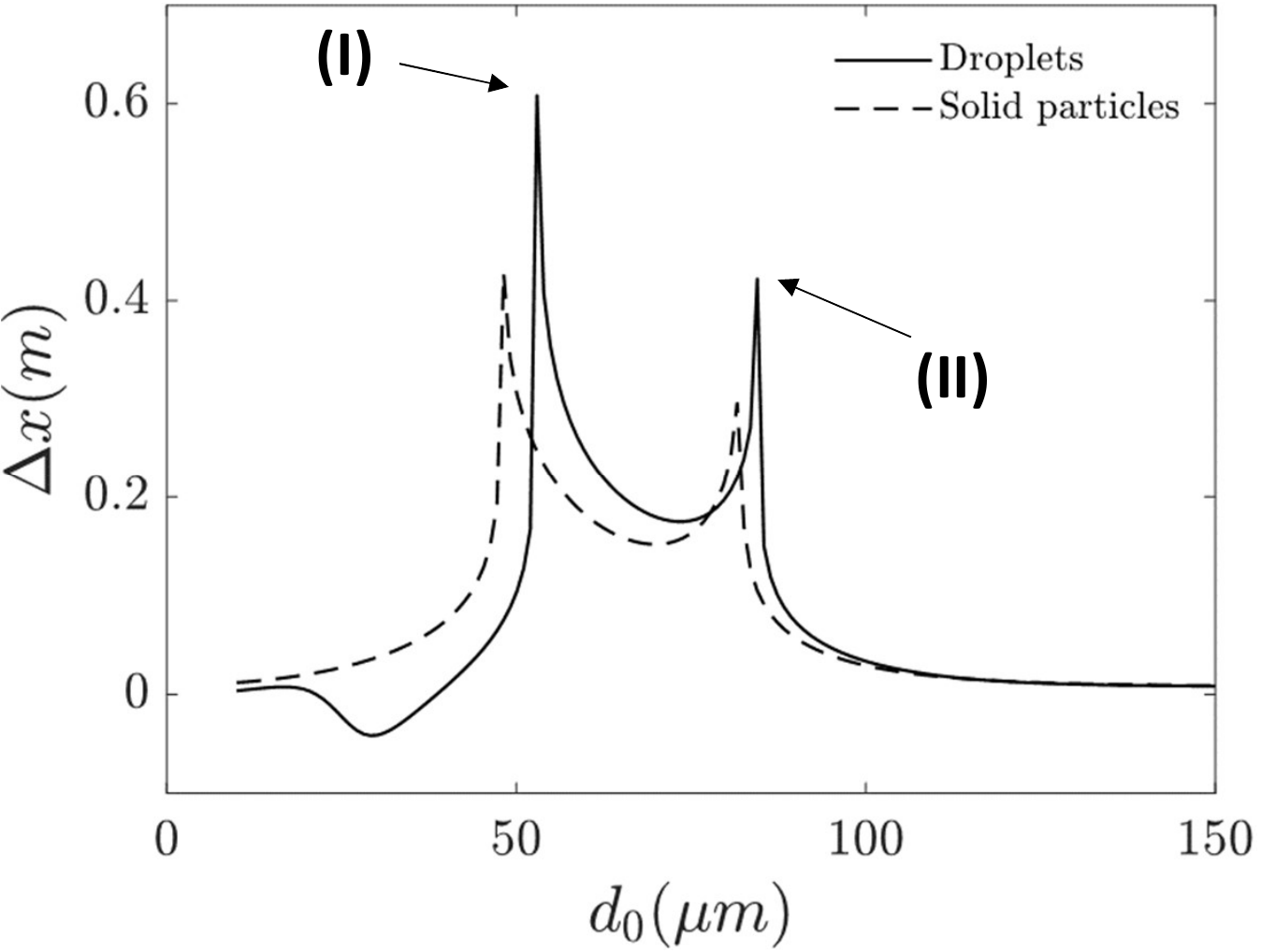}}
  \caption{Downstream settling distances for droplets and particles of variable initial diameter, dispersed by a vortex of $a=0.1m$ diameter and $\omega _{m} =25 \hspace{0.1cm} 1/s$ maximal intensity. Both were released from $(-2,1)$ relative to the vortex center.}
  \label{fig:d0_delX}
\end{figure}

We now seek to isolate and analyze the dependence of the settling distance on the initial diameter of the droplets and the wake structure intensity.
Ergo, we explore the settling distances of droplets and particles released at the same location $(-2,1)$ in two cases; for constant maximal vorticity of $\omega _{m} =25 \hspace{0.1cm} 1/s$ and varying initial diameter, as well as for constant initial diameter $d_0=65 \mu m$, and vortices of different maximal vorticities.

The impact of the evaporation process is conspicuous when comparing the results obtained in Fig.~\ref{fig:d0_delX} for small diameters. As previously discussed, small ($d_0<40\mu m$ droplets are pushed oppositely to the vortex direction, whereas solid particles can translate only downstream.
On the other hand, large ($d_0>100\mu m$ droplets and particles have similar behavior. The two were not entrained significantly by the dipole due to their large initial masses and insignificant evaporation.

Notably, droplets in the intermediate range might be influenced by the vortex momentously. Two sharp peaks in the settling distance are observed for droplets around $d_0=65\mu m$ $d_0=80\mu m$. This "double critical" result suggests that the dispersion of the droplet by a wake structure is highly sensitive to the diameter of the dispersed droplets, as slight changes in the diameter may lead to contrasting results.
As for solid particles, a comparable trend was found. It should be noted that the maximal displacement of particles is smaller compared to the displacement of the droplet, as evaporating droplets lose mass, thus allowing for the greater residence time inside the vortex.
Furthermore, the lack of evaporation causes the "critical" diameters of the particles to transpose towards lower values. This, in turn, leads to significant, alternating disparities between them; while a $50\mu m$ particle is carried $0.4m$ downstream, a droplet will translate only $0.05m$.
Contrarily, a $80\mu m$ droplet is carried $0.4m$, whereas a particle of the same size and location will travel only $0.06m$.

\begin{figure}
\centerline{\includegraphics[width=0.65\linewidth]{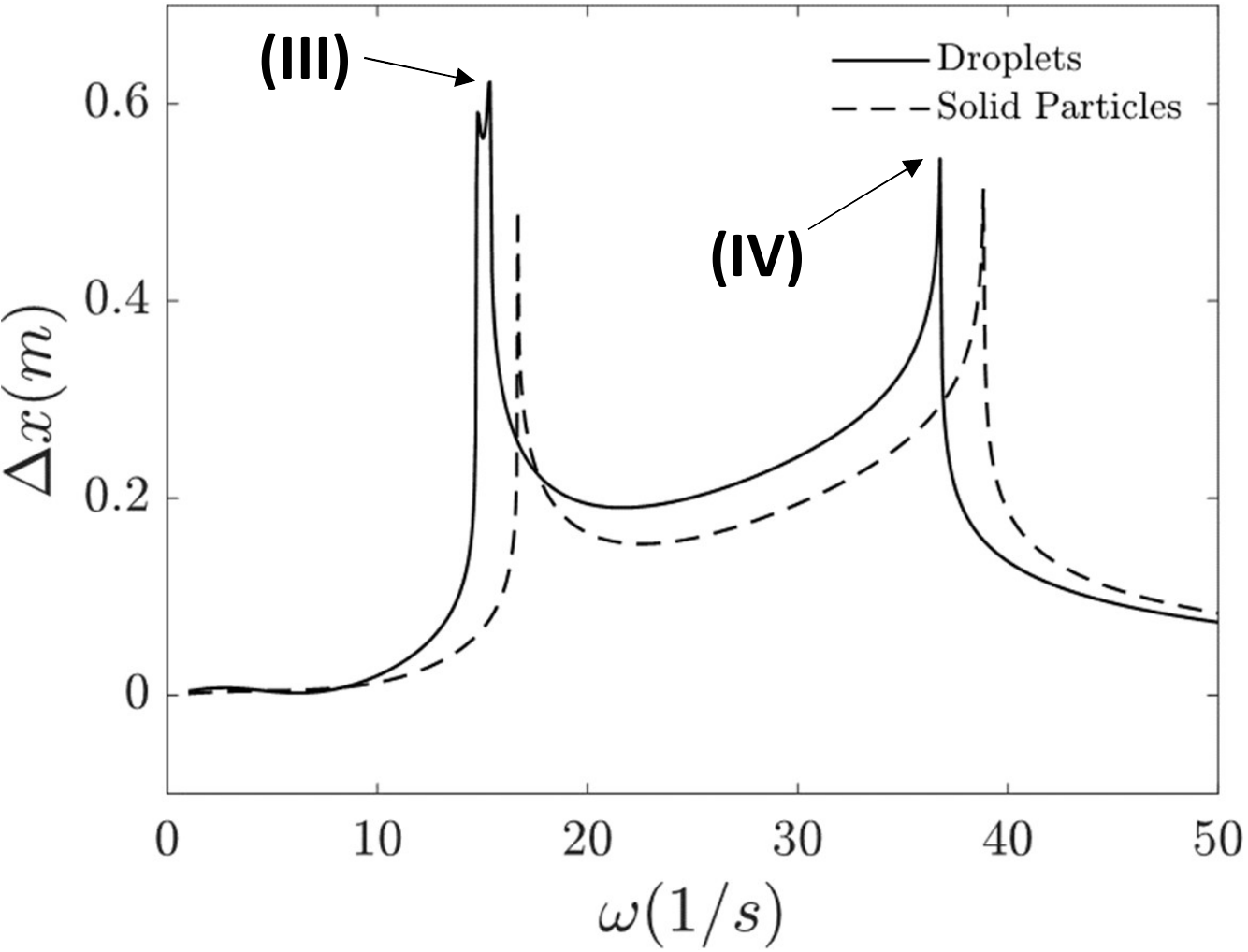}}
    \caption{Downstream settling distances of a droplet and a particle, dispersed by vortices of $a=0.1m$ diameter and variable maximal intensities $\omega _{m}$. The particle and droplet initial diameter is $d_0=65 \mu m$, and both were released from $(-2,1)$ relative to the vortex center.}
    \label{fig:w_delX}
\end{figure}

The settling distance and airborne time dependence on the wake structure intensity are clearly shown in Fig.\ref{fig:w_delX}. As expected, low-intensity vortices do not affect the droplets significantly, as the flow fields they induce are not strong enough to overcome the droplet inertia. 
However, at the other limit of high-intensity, vortices did not scatter the droplet farthest, as perhaps expected. Although such dipoles might be more energetic, their translation velocities $U_0$ are, accordingly, higher. As a result, the strong vortices pass the vicinity of the droplet rapidly, not allowing for the particles and droplet to entrap inside the core. Thus, overall, increasing the vorticity might decrease the settling distance of the droplets. 
As in Fig.~\ref{fig:d0_delX}, the intermediate region is characterized by a "double critical" behavior exhibited by both droplets and particles. The difference between droplets and particles settling distances is evident as the droplets respond to dipoles of lower vorticity, and the maximal displacement of particles is smaller compared to the displacement of the droplet. The droplet's evaporation causes dissimilarity, as evaporation lowers the droplet's relaxation time, requiring less powerful vortices in order to entrain the droplet. As a result, the critical vorticities for droplets were shifted to lower values compared to particles.

\begin{figure}
\centerline{\includegraphics[width=0.65\linewidth]{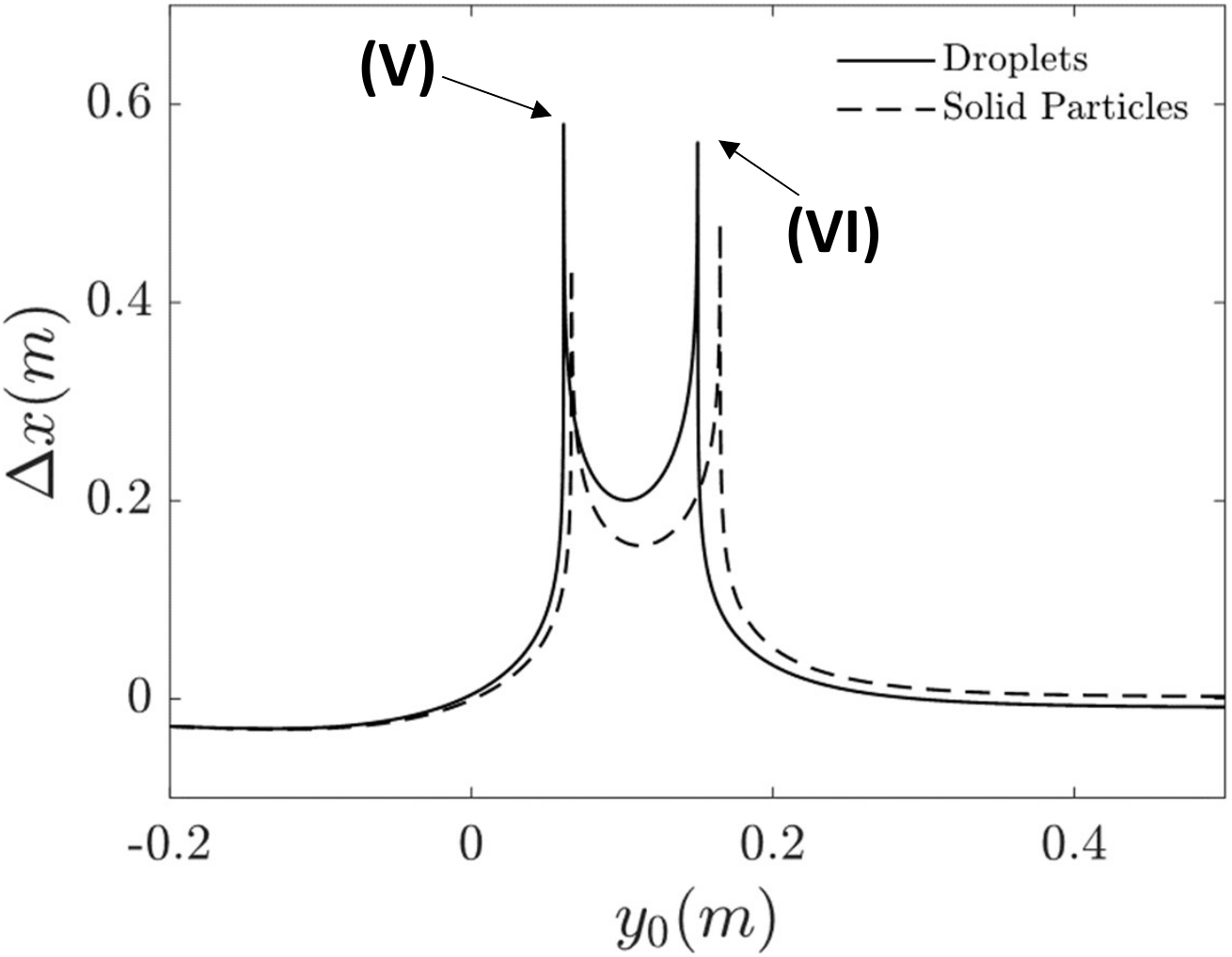}}
    \caption{
    Downstream settling distances of droplets and particles dispersed by a vortex of $a=0.1m$ diameter and $\omega _{m} =25\hspace{0.1cm}1/s$ maximal intensity. Both were released from $x=-2a$ and variable vertical distances relative to the vortex center. The particle and droplet initial diameter is $d_0=65 \mu m$}
    \label{fig:y0_delX}
\end{figure}

The droplet's initial location was also investigated, as demonstrated in Fig.\ref{fig:y0_delX}. We note that both droplets and particles placed below the center ($y_0<0$) were driven backward, as expected. The relation between the initial location and the settling distance is showing the same trend as Fig.\ref{fig:d0_delX} and Fig.\ref{fig:w_delX}, further revealing the unique scattering patterns.

\begin{figure*}
    \centering
    \centerline{\includegraphics[width=1\textwidth]{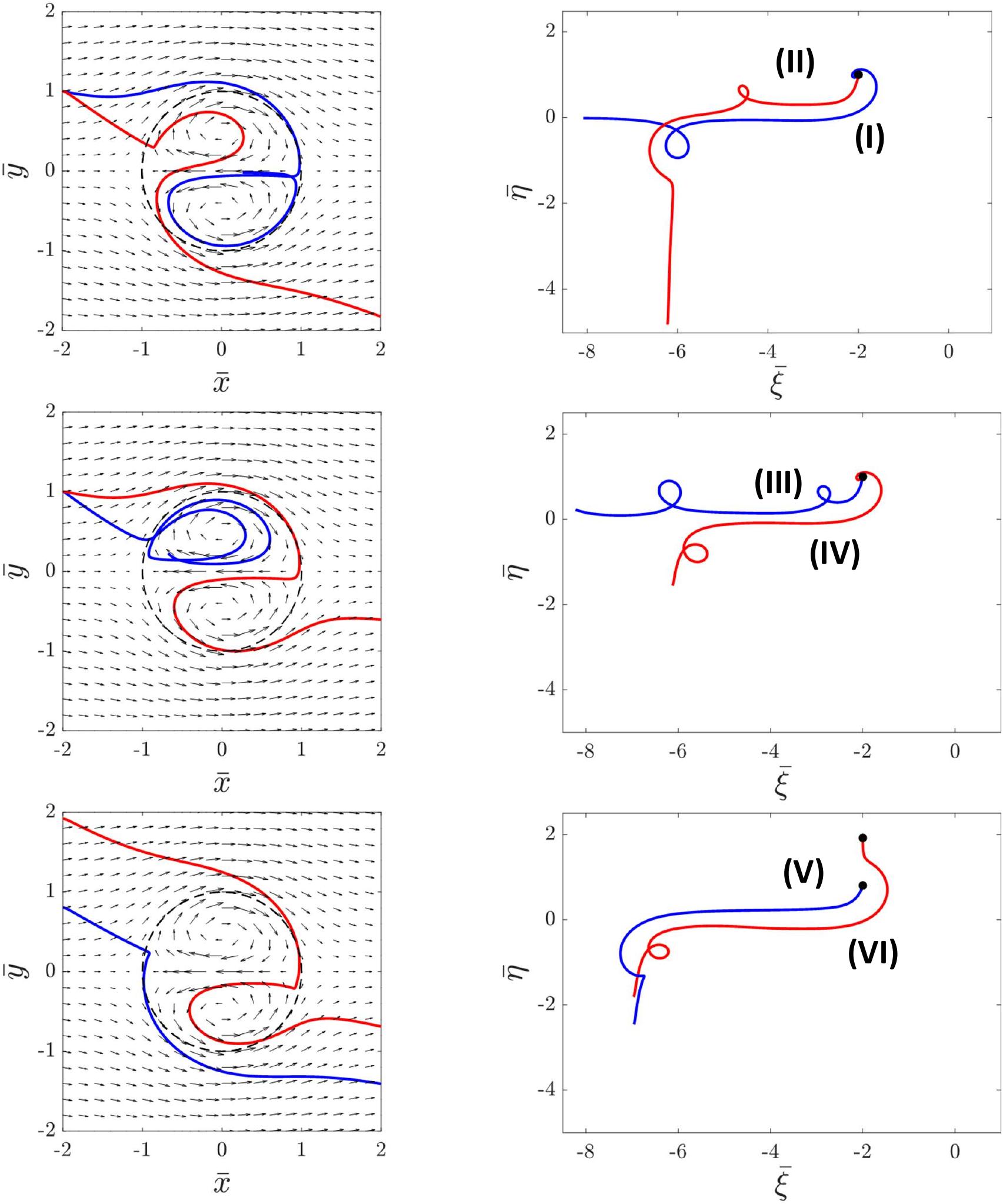}}
    \caption{Comparison of evaporating droplet's critical trajectories.}
    \label{fig:critical}
\end{figure*}

The droplet's critical trajectories (labeled I-VI in Fig. \ref{fig:d0_delX}-\ref{fig:y0_delX} and Table \ref{tab}) are presented in Fig\ref{fig:critical}, in both stationary (left subplots) and a frame of reference following the vortex center (right subplots).
\begin{table}[h]
\centering
    \begin{tabular}{c@{\hspace{0.5in}}c@{\hspace{0.2in}}c@{\hspace{0.2in}}c}
    Trajectory & $d_0~(\mu m)$ &$\omega_m~(1/s) $& $y_0~(m)$ \\ \hline
    I & 52 & 25 & 0.1 \\   
    II & 85 & 25 & 0.1 \\ 
     III & 65 & 15 & 0.1 \\ 
      IV & 65 & 36 & 0.1 \\ 
       V & 65 & 25 & 0.06 \\ 
        VI & 65 & 25 & 0.15 \\ \hline
    \end{tabular}
    \caption{List of critical trajectories.}
    \label{tab}
\end{table}
Curiously, although significant differences exist between the trajectories, they all penetrate the vortex core at the same locations- either near the left  (II, III, and V) or the right (I, IV, and VI) stagnation point. This result might clarify the origin of the "double critical" relations abovementioned.
As discussed, the passing of the droplet near the stagnation point enhances its settling distance significantly, as the droplet translates downstream along with the vortex.
Thus, we may identify two critical trajectories, one for each stagnation point. Slight changes in the droplet diameter, initial location, or vortex intensity might sway the droplet away from the point, hence significantly reducing its entrainment by the vortex.

To highlight this realization, we shall investigate trajectory I.
Since the droplet has penetrated the core at the back stagnation point, we may conclude that a lighter droplet would have washed downstream and thus influenced by the vortex only briefly.
Following the same logic, a heavier droplet would have entered the core earlier and closer to the vortex maximal vorticity point.
Consequently, it would not have crossed the vicinity of the stagnation point but rather swirled inside the core and ejected rapidly, leading to a sharp decrease of the settling distance.
Further increase of the droplet mass will move the penetration location closer to the front stagnation point (trajectory II), and as a result, leads to a sharp increase in the droplet airborne time.
We may apply the same analysis for the rest of the results, therefore offering a general interpretation of the findings presented in Fig. \ref{fig:d0_delX}-\ref{fig:y0_delX}.

Trajectories I and III demonstrate that droplets might get caught and evaporate inside the vortex.
While the droplets had sufficient inertia to enter the core, they lost mass due to evaporation while rotating inside the vortex core, and as a result, could not escape.
This insight might be of significance when investigating the dynamics of aerosols, particularly those that originated from respiratory droplet evaporation.
In this case, some aerosols might appear within the vortex core and may be carried along with the vortex downstream, leading to substantial and unanticipated dispersion.

%SECTION4
\section{\label{sec:conc} Conclusions}

A mathematical analysis for the dynamics and dispersion of droplets and particles under the influence of an approaching dipolar vortex was carried out.
Our analysis reveals the complex dynamic interactions of evaporating droplet dynamics and the induced wake flow, affecting the settling distances of micron-sized droplets.

Droplets that pass close to the vortex core stagnation points are captured and translated by the vortex into farther distances. Such droplets exhibited the largest settling distances, up to six times the vortex core length scale.
Other droplets that do not pass in the proximity of the stagnation points exhibit significantly shorter settling distances.
The large disparity of droplet trajectory was observed for droplets of either larger or smaller initial diameters, dispersed by a vortex of different intensity, or droplets initially positioned nearby.

Particular attention was given to the coupled aerodynamic and thermodynamic properties of an individual droplet.
The effect of evaporation was analyzed and studied by comparing the trajectories and settling distances of the evaporating droplet and a solid particle. We found that the settling distances due to the vortex dispersion effect cannot be determined for a specific initial droplet due to the highly sensitive dynamic interaction between the flow and the droplets.

The model proposed for the dynamic interaction might offer a new analytic tool for the assessment of the spreading mechanisms of airborne-carried pathogens, such as SARS-CoV-2.
Unlike other works in the field, the presented model assumes the respiratory droplets are suspended in undisturbed air and thus might allow for a better assessment of the possible dispersion of virus-loaded droplets after an exhalation event. 
Taking an analytic approach has allowed for the isolation of the interaction between a droplet and the vortical flow, and thus may aid in uncovering the basic principles behind the reported enhanced entrainment of droplets by characteristic vortical flows.
Our results suggest that micron-sized droplets can spread significantly in the presence of wake flow structures, constantly generated indoors due to a movement of objects or flow.

The extension of our approach in rationalizing more complex droplet dynamics is currently being examined. We aim to incorporate more complex heat and mass convection models into our model, consider aerosol formation mechanisms, account for the different chemical compositions of the droplet, and investigate the process for changeable environmental conditions. Moreover, the enhanced risk for infection due to the wake flow can be evaluated by considering the actual viral load carried by the entrained droplets, and thus offer a better assessment of the importance of the presented phenomenon during the current COVID-19 outbreak.

\section*{Acknowledgement}
This research was supported by the ISRAEL SCIENCE FOUNDATION (grant No. 1762/20).

\singlespacing
\bibliographystyle{elsarticle-num-names}
\bibliography{references}

\end{document}